# NEW LINEAR THEORY OF HYDRODYNAMIC INSTABILITY OF THE HAGEN-POISEUILLE FLOW

S.G. Chefranov[1], A.G. Chefranov[2]

[1]A.M. Obukhov Institute of Atmospheric Physics RAS, Moscow, Russia; schefranov@mail.ru

[2]Eastern Mediterranean University, Famagusta, North Cyprus; Alexander.chefranov@emu.edu.tr

Abstract

New condition $\text{Re} > \text{Re}_{th_{\min}} \approx 124$ of linear (exponential) instability of the Hagen-Poisseuille (HP) with respect to extremely small by magnitude axially-symmetric disturbances of the tangential component of the velocity field is obtained. For this, disturbances must necessarily have quasi-periodic longitude variability (not representable as a Fourier series or integral) along the pipe axis that complies with experimental data and differs from the usually considered idealized case of pure periodic disturbances for which HP flow is stable for arbitrary large Reynolds numbers Re. Obtained minimal threshold Reynolds number is related to the spatial structure of disturbances (having two radial modes with non-commensurable longitudinal periods) in which irrational value $p \approx 1.58..$ of the ratio of the two longitudinal periods is close to the value of the "golden ratio" equal to 1.618…

# NEW LINEAR THEORY OF HYDRODYNAMIC INSTABILITY OF THE HAGEN-POISEUILLE FLOW

S.G. Chefranov[1)], A.G. Chefranov[2)]

[1)] A.M. Obukhov Institute of Atmospheric Physics RAS, Moscow, Russia; schefranov@mail.ru

[2)] Eastern Mediterranean University, Famagusta, North Cyprus; Alexander.chefranov@emu.edu.tr

1. Problem of defining mechanisms of hydrodynamic instability leading to the turbulization of the Hagen-Poiseuille (HP) flow has fundamental and application worth.[1]

So, currently, it is decided [1-4] that HP flow is exponentially stable with respect to extremely small by magnitude disturbances for any large Reynolds number $\text{Re} = \frac{V_{\max} R}{\nu}$, where $V_{\max}$ - maximal (near axis) velocity of the HP flow in the pipe of radius $R$, and $\nu$ - coefficient of kinematic viscosity. Such a conclusion of the linear theory of hydrodynamic stability is based on the traditional consideration of pure periodic disturbances when different radial modes have the same period of longitudinal variability along the pipe axis.

In [5], it is shown that conclusion about linear stability of HP flow needs clarification since if instead of periodic to consider conditionally-periodic (quasi-periodic) disturbances, then already for finite $\text{Re}$ it might happen linear (exponential, not algebraic [6, 7]) instability of HP flow.

In the present paper, we further develop representation [5] in the frame of the new theory of linear instability of HP flow. Meanwhile, contrary to [5] in particular we show possibility of getting a threshold by $\text{Re}$ condition of linear instability of HP flow which does not depend on the procedure of averaging when using Galerkin approximation (necessary because of consideration of longitudinal quasi-periodicity of disturbances resulting from different longitudinal periods of different radial modes).

2. Let's consider evolution in time of axially symmetric extremely small by magnitude hydrodynamic disturbances of the tangential component of the velocity field $V_\varphi$ in the cylindrical system of coordinates $(z, r, \varphi)$:

$$\frac{\partial V_\varphi}{\partial t} + V_{0z}(r) \frac{\partial V_\varphi}{\partial z} = \nu (\Delta V_\varphi - \frac{V_\varphi}{r^2}), \tag{1}$$

---

[1] HP flow – laminar static flow of uniform viscous non-compressible fluid along static straight and infinite by length pipe with circular cross section.

where $V_{0z}(r) = V_{max}(1 - \frac{r^2}{R^2})$, $V_{0r} = V_{0\varphi} = 0$ - main (disturbed) HP flow along the pipe of radius $R$ having in it a constant longitudinal pressure gradient $\frac{\partial p}{\partial z} = const$, when $V_{max} = \frac{R^2}{4\rho\nu} \cdot \frac{\partial p_0}{\partial z}$ with constant density $\rho$ of the uniform fluid. In (1), $\Delta$ is Laplace operator. Due to assumed axial symmetry of the extremely small disturbances $V_\varphi$ ( i.e. since $\frac{\partial V_\varphi}{\partial \varphi} = 0$), in the right-hand side of (1), there is no derivative $\frac{\partial p}{\partial \varphi}$ from small disturbance of the pressure field $p$. Meanwhile, (1) allows closed description of evolution of pure tangential disturbances of HP flow.

Let's find solution of equation (1) in the following form

$$V_\varphi = V_{max} \sum_{n=1}^{N} A_n(z,t) J_1(\gamma_{1,n} \frac{r}{R}), \qquad (2)$$

which automatically meets boundary conditions of finiteness of $V_\varphi$ for $r = 0$ and non-slipping $V_\varphi = 0$ for $r = R$ on the hard pipe boundary since $J_1$ is the Bessel function of the first order, and $\gamma_{1,n}$ are zeroes of that function ($n = 1, 2, ..$).

Using feature of orthonormality of Bessel functions and a standard averaging procedure in the Galerkin approximation (see [2]), one gets in the dimensionless form from (1), (2) the following closed system of equations for the functions $A_n(z,t)$:

$$\frac{\partial A_m}{\partial \tau} + \gamma_{1,m}^2 A_m - \frac{\partial^2 A_m}{\partial x^2} + \text{Re} \sum_{n=1}^{N} p_{nm} \frac{\partial A_n}{\partial x} = 0. \qquad (3)$$

In (3), $x = \frac{z}{R}$, $\tau = \frac{t\nu}{R^2}$, $m = 1, 2, ..., N$, and coefficients $p_{nm}$ are as follows:

$$p_{nm} = \delta_{nm} - \frac{2}{J_2^2(\gamma_{1,m})} \int_0^1 dy\, y^3 J_1(\gamma_{1,n} y) J_1(\gamma_{1,m} y), \qquad (4)$$

where $J_2$ is a Bessel function of the second order, , $\delta_{nm}$ is Kronekker's symbol ($\delta_{nm} = 1$ for $n = m$ and $\delta_{nm} = 0$ if $n \neq m$). Obviously that $p_{12} \neq p_{21}$ in (4) due to the presence of a factor before the integral in (4) (since $J_2(\gamma_{1,1}) \neq J_2(\gamma_{1,2})$).

Let's limit ourselves by the case of $N = 2$. Let amplitudes $A_1$ and $A_2$ corresponding to different modes of radial variability $V_\varphi$ in (2) have different periods of variability along the pipe axis, i.e. let's consider $A_1$ and $A_2$ as follows

$$A_1(x,t) = A_{10}e^{\lambda\tau+i2\pi\alpha x}, \quad A_2(x,t) = A_{20}e^{\lambda\tau+i2\pi\beta x} \tag{5}$$

where $\alpha \neq \beta$ contrary to the usual (see [2]) consideration of the problem of stability of the HP flow in the linear approximation by amplitudes of disturbances. Meanwhile, the value of $\lambda = \lambda_1 + i\lambda_2$ in (5) defines the same (synchronous) character of dependency of functions $A_1$ and $A_2$ on time. Substitution of (5) in (3) (for $N = 2$) leads to the following system

$$(\lambda + \gamma_{1,1}^2 + 4\pi^2\alpha^2 + i\operatorname{Re} p_{11} \cdot 2\pi\alpha)A_{10}e^{i2\pi\alpha x} + i\operatorname{Re} p_{21} 2\pi\beta A_{20}e^{i2\pi\beta x} = 0 \tag{6}$$

$$i\operatorname{Re} p_{12} 2\pi\alpha A_{10}e^{i2\pi\alpha x} + (\lambda + \gamma_{1,2}^2 + 4\pi^2\beta^2 + i\operatorname{Re} p_{22} \cdot 2\pi\beta)A_{20}e^{i2\pi\beta x} = 0 \tag{7}$$

System (6), (7) admits exact solution for constant coefficients $A_{10}$ and $A_{20}$ only in the case when $\alpha = \beta$ and in (5) functions $A_1$ and $A_2$ have the same pure periodic character of variability along the pipe axis. It is not difficult to check that from the condition of solvability of uniform system (6), (7) there may be obtained well-known conclusion [1-4] about linear stability of HP flow since in that case it is found out that for any $\operatorname{Re}$, $\lambda_1 < 0$.

3. Considering quasi-periodic variability of $V_\varphi$ along the pipe axis for $\alpha \neq \beta$ in (5), we use Galerkin approximation to solve the system (6), (7). Meanwhile, let's average (6) multiplying (6) by the function $e^{-i2\pi\gamma_1 x}$ and integrating over $x$ in the limits from 0 to $1/\gamma_1$ (i.e. applying to (6) an operation of $|\gamma_1| \int_0^{1/|\gamma_1|} dx$, where $|\gamma_1|$ is the modulus of $\gamma_1$). The equation (7) is averaged applying to (7) the same as in (6) operation of averaging but with replaced in it $\gamma_1$ by $\gamma_2$, where in the general case $\gamma_1 \neq \gamma_2$.

Solvability condition of the system of equations obtained from (6), (7) after the specified above averaging is the following dispersion equation for $\lambda$:

$$\lambda^2 + \lambda(a+b) + ab + gc = 0, \tag{8}$$

where complex value $\lambda = \lambda_1 + i\lambda_2$ is uniquely defined by the following coefficients:

$a = \gamma_{1,1}^2 + 4\pi^2\alpha^2 + i\operatorname{Re}2\pi\alpha p_{11} \equiv a_1 + ia_2$, $b = \gamma_{1,2}^2 + 4\pi^2\beta^2 + i\operatorname{Re}2\pi\beta p_{22} \equiv b_1 + ib_2$,

$c = 4\pi^2\alpha\beta\operatorname{Re}^2 p_{12}p_{21}$ and $g = \dfrac{I_{2\alpha}I_{1\beta}}{I_{2\beta}I_{1\alpha}}$.

Here, in g, we have values of elementary integrals: $I_{m\alpha} = |\gamma_m|\int_0^{1/|\gamma_m|}dxe^{i2\pi x(\alpha-\gamma_m)}$ and $I_{m\beta} = |\gamma_m|\int_0^{1/|\gamma_m|}dxe^{i2\pi x(\beta-\gamma_m)}$, where $m=1,2$. From (8), one can obtain (see[5]) the condition of the linear instability of the HP flow when in (8) $\lambda_1 > 0$ for some $\operatorname{Re} > \operatorname{Re}_{th}$. Result in that case is significantly depending on the value of $g$, defined by the way of averaging of the system (6), (7) on the base of Galerkin approximation.

So, if we change the averaging procedure (applying operation $|\gamma_1|\int_0^{1/|\gamma_1|}d\chi e^{-i2\pi\gamma_1\chi}$ already to the equation (7), not to (6), as it was done above; and vice versa, we apply to (6) the operation of averaging applied above for averaging of the equation (7)), then, in the dispersion equation, value of $g$ is replaced by $1/g$.

We require that the conclusion on the stability of HP flow should not depend on the pointed difference in the averaging procedure conducting that is possible only when $g^2 = 1$. This equation for $g$ has two roots, $g = 1$, and $g = -1$. For $g = 1$, conclusion on the stability of HP flow exactly coincides with the case of pure periodic disturbances when $\alpha = \beta$.

Let's consider the second case of $g = -1$, and show that meanwhile linear instability of HP flow is possible already for finite value of the threshold Reynolds number $\operatorname{Re}_{th}$.

Actually, for $g = -1$ from (8) it follows that $\lambda_1 > 0$ when

$$\operatorname{Re} > \operatorname{Re}_{th} = \frac{F}{2\pi\sqrt{D}}, \qquad (9)$$

where $F = (a_1 + b_1)\sqrt{a_1b_1}$, $D = (a_1 + b_1)^2\alpha\beta p_{12}p_{21} - a_1b_1(p_{11}\alpha - p_{22}\beta)^2$.

Obviously, for realization of linear exponential instability of HP flow $D > 0$ is necessary, that is trivially to be met when $(p_{11}\alpha - p_{22}\beta)^2 < 4\alpha\beta p_{12}p_{21}$.

If to introduce a parameter $p = \dfrac{\alpha}{\beta}$, defining ratio of periods in (5), then from the pointed inequality providing positiveness of $D$ in (9), it follows that the following holds

$$x_- < p < x_+, \qquad (10)$$

where $x_\pm = \dfrac{p_{11}p_{22} + 2p_{12}p_{21} \pm 2\sqrt{p_{12}p_{21}(p_{11}p_{22} + p_{21}p_{12})}}{p_{11}^2}$, i.e. according to (4), $x_+ = 1.739..., x_- = 0.588...$

From the other hand, from the condition $g = -1$ it follows that the following inequality holds

$$g = \frac{(1 - e^{i\frac{2\pi\beta}{|\gamma_1|}})(1 - e^{i\frac{2\pi\alpha}{|\gamma_2|}})(\alpha - \gamma_1)(\beta - \gamma_2)}{(1 - e^{i\frac{2\pi\beta}{|\gamma_2|}})(1 - e^{i\frac{2\pi\alpha}{|\gamma_1|}})(\beta - \gamma_1)(\alpha - \gamma_2)} = -1. \qquad (11)$$

In particular, equation (11) is satisfied when $\beta = \alpha + |\gamma_1|n$ and $\beta = \alpha + |\gamma_2|m$, where $m, n$ - are any integers having the same sign since with necessity then holds $\dfrac{|\gamma_1|}{|\gamma_2|} = \dfrac{n}{m} > 0$. Meanwhile from (11), it follows that the following relation defining the value $p = \dfrac{\alpha}{\beta}$ depending on the values of $m, n$ and signs of $\gamma_1, \gamma_2$ holds:

$$p = \frac{1}{1 + \dfrac{2}{B}}, \quad B = B_\pm = \frac{\gamma_1}{n|\gamma_1|} + \frac{\gamma_2}{m|\gamma_2|} - 1 \pm \sqrt{1 + \frac{1}{m^2} + \frac{1}{n^2}}. \qquad (12)$$

Obviously, $p$ from (12) shall meet inequality (10). In particular, for $m = n = 1$ value of $p$ (when $B = B_-$ and $\gamma_1 > 0, \gamma_2 > 0$) is $p \approx 1.58...$ Since $\text{Re}_{th}$ in (9) is a function of $\beta$ and $p$, for the pointed value of $p$, meeting inequality (10), from (9), we can get that minimal value $\text{Re}_{th\min} \approx 124$ is reached in the proximity of $\beta \approx 0.5$.

Thus, it is found out possibility of linear (exponential) instability of HP flow already for $\text{Re} > \text{Re}_{th\min} \approx 124$, that does not contradict to the well-known estimates of guaranteed stability of HP flow obtained from energy considerations (see [1]) for $\text{Re} < 81$. Obviously, exponential growth of $V_\varphi$ after reaching of some finite values shall be replaced be a new non-linear mode of evolution in which all components of velocity and pressure are already all cross-linked.

Thus far, in the present work, new conclusion on linear instability of HP flow is obtained. It could not be obtained earlier because in the systems of type (3) always traditionally only the case of very the same periodic (or quasi-periodic that does matter) longitudinal variability is considered for all functions in (3) for different m=1…N.